\documentclass[twocolumn]{emulateapj}

\usepackage{color}
\usepackage{natbib}
\usepackage[colorlinks, allcolors=blue]{hyperref}
\newcommand{\arcsecs}{\hbox{$^{\prime\prime}$}}

\usepackage{graphicx}
\usepackage{txfonts}
\usepackage{microtype}

\begin{document}

\title{Spatially Resolved Signatures Of Bi-Directional Flows Observed\\ In Inverted-Y Shaped Jets}
\author{C. J. Nelson}
\affil{Astrophysics Research Centre (ARC), School of Mathematics and Physics, Queen’s University, Belfast, BT7 1NN, Northern Ireland, UK}
\affil{Solar Physics and Space Plasma Research Centre (SP2RC), School of Mathematics and Statistics, University of Sheffield, Hicks Building, S3 7RH, UK}

\author{N. Freij}
\affil{Departament de F\'isica, Universitat de les Illes Balears, 07122 Palma de Mallorca, Spain}

\author{S. Bennett}
\affil{Solar Physics and Space Plasma Research Centre (SP2RC), School of Mathematics and Statistics, University of Sheffield, Hicks Building, S3 7RH, UK}

\author{R. Erd{\'e}lyi}
\affil{Solar Physics and Space Plasma Research Centre (SP2RC), School of Mathematics and Statistics, University of Sheffield, Hicks Building, S3 7RH, UK}
\affil{Department of Astronomy, E\"otv\"os Lor\'and University, P\'azm\'any P\'eter s\'et\'any 1/A, H-1117 Budapest, Hungary}

\author{M. Mathioudakis}
\affil{Astrophysics Research Centre (ARC), School of Mathematics and Physics, Queen’s University, Belfast, BT7 1NN, Northern Ireland, UK}

\shortauthors{Nelson et al.}
\shorttitle{Bi-Directional Flows In Inverted-Y Shaped Jets At The Foot-Points Of Surges}
\email{c.j.nelson@sheffield.ac.uk}

\begin{abstract}
Numerous apparent signatures of magnetic reconnection have been reported in the solar photosphere, including  inverted-Y shaped jets. The reconnection at these sites is expected to cause localised bi-directional flows and extended shock waves; however, these signatures are rarely observed as extremely high spatial-resolution data are required. Here, we use H$\alpha$ imaging data sampled by the Swedish Solar Telescope's CRisp Imaging SpectroPolarimeter to investigate whether bi-directional flows can be detected within inverted-Y shaped jets near the solar limb. These jets are apparent in the H$\alpha$ line wings, while no signature of either jet is observed in the H$\alpha$ line core, implying reconnection took place below the chromospheric canopy. Asymmetries in the H$\alpha$ line profiles along the legs of the jets indicate the presence of bi-directional flows, consistent with cartoon models of reconnection in chromospheric anemone jets. These asymmetries are present for over two minutes, longer than the lifetimes of Rapid Blue Excursions, and beyond $\pm1$ \AA\ into the wings of the line indicating that flows within the inverted-Y shaped jets are responsible for the imbalance in the profiles, rather than motions in the foreground. Additionally, surges form following the occurrence of the inverted-Y shaped jets. This surge formation is consistent with models which suggest such events could be caused by the propagation of shock waves from reconnection sites in the photosphere to the upper atmosphere. Overall, our results provide evidence that magnetic reconnection in the photosphere can cause bi-directional flows within inverted-Y shaped jets and could be the driver of surges.
\end{abstract}

\keywords{Sun: activity - Sun: photosphere - Sun: chromosphere}

\section{Introduction}
\label{Introduction}

A huge range of potential signatures of magnetic reconnection in the solar photosphere have been reported in the literature. Transient, small-scale inverted-Y shaped jets, for example, have been widely studied over the past decade using high-spatial and temporal resolution data. Such inverted-Y shaped jet events were first identified by \citealt{Shibata07} and manifest as bright regions on broad-band \ion{Ca}{2} H images (\citealt{Morita10, Nishizuka11, Singh12}), as well as other chromospheric lines. Cartoon models presented in the literature (see, for example, \citealt{Shibata07, Singh11}) suggest that bi-directional reconnection out-flows should exist in one foot-point of the inverted-Y shaped jets; however, hints that such bi-directional flows exist have only rarely been reported to date (see, for example, \citealt{Zeng16, Tian18}) as wide-band imaging data have typically been used to study these events. Recently, similar events have been identified at the foot-points of coronal loops (\citealt{Chitta17}) and in sunspot light-bridges (\citealt{Tian18}) indicating that inverted-Y shaped jets and, hence, magnetic reconnection may be prevalent throughout the solar photosphere. It is, therefore, important that further analysis of these events is conducted using high-spatial and temporal resolution data in order to better understand their formation and evolution. We provide some steps in this direction in this article.

\begin{figure*}
\includegraphics[width=0.95\textwidth]{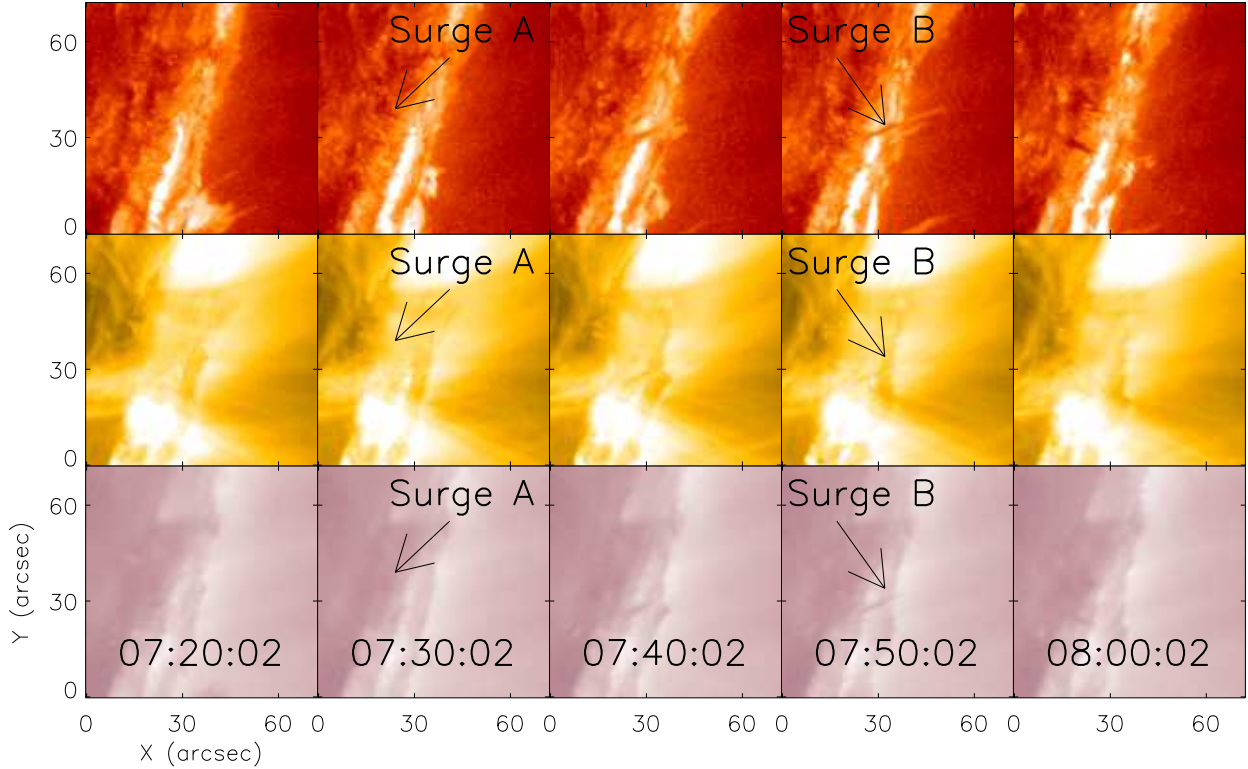}
\caption{A time-series of SDO/AIA context images displaying the evolution of the FOV studied here with the intensity log-scaled to enhance the contrast. The top, middle, and bottom rows plot data sampled by the $304$ \AA, $171$ \AA, and $211$ \AA\ filters, respectively. The locations of the surges analysed later in this article (`Surge A' itself is not easily observed in these zoomed-out context images due to its relatively small size), which appear as absorption features in each of the SDO/AIA channels, are indicated by the appropriately labelled white arrows in the top ($304$ \AA) row.}
\label{SDOtime}
\end{figure*}

Further examples of reported reconnection events in the lower solar atmosphere are Ellerman bombs (EBs; \citealt{Ellerman17}) and Quiet-Sun Ellerman-like Brightenings (QSEBs; \citealt{Rouppe16, Nelson17}). These events appear as small-scale (diameters often less than $1$\arcsecs), short-lived (lifetimes of less than $10$ minutes) brightenings in the wings of the H$\alpha$ line, often co-spatial to regions of cancelling magnetic flux (see, for example, \citealt{Reid16}). Simulations of EBs have strongly supported the idea that such events form as a response to magnetic reconnection in the photosphere (see, for example, \citealt{Nelson13b, Danilovic17, Hansteen17}). 

Interestingly, both inverted-Y shaped jets and EBs have been reported to occur at the foot-points of larger-scale chromospheric events such as chromospheric anemone jets and surges (e.g., \citealt{Roy73, Watanabe11, Heesu13}). Surges are columns of relatively cool material extending out from the solar chromosphere into the corona. These events typically have lengths of around $10$-$70$\arcsecs\ (\citealt{Roy73b}) and have been shown to consist of numerous distinct thin threads of material (\citealt{Nelson13, Li16}). Surges can be identified in chromospheric lines observed from the ground or in transition region and coronal lines sampled by satellites and often occur during interactions (e.g., through flux cancellation) between newly emerged magnetic flux and the background photospheric magnetic field (\citealt{Chae99, Guglielmino10, Nobrega16}). As such, surges are considered to be formed as a response to magnetic reconnection in the lower solar atmosphere (for example, \citealt{Roy73, Guglielmino10, Nobrega16}). 

It should be noted, however, that only a small minority of the EBs reported in the literature appear to occur co-spatial to these large chromospheric ejections. Indeed, \citet{Watanabe11} found only $2$ of the $17$ EBs they studied formed co-spatial to surges. More recently, \citet{Reid15} presented an observation showing short jets (not directly identified as surges by those authors) appearing to be driven by an EB observed in the H$\alpha$ line wings. The reason why only a small percentage of reconnection events in the photosphere drive surges is currently unknown, however, it could speculatively be due to the local magnetic field topologies, the height of the reconnection, or the reconnection rate in the photosphere. This remains to be seen in future research using up-coming telescopes such as DKIST.

Although surges are suggested to be driven by magnetic reconnection in the lower solar atmosphere, they are not thought to be actual reconnection out-flows. It has been hypothesised that slow magnetohydrodynamic (MHD) wave pulses are excited at the reconnection site and that these waves propagate up through the atmosphere to the transition region where they form shocks. These shocks create pressure gradients which drag the relatively dense material contained at the transition region to greater heights in the solar atmosphere, thereby creating the observed signatures of surges. This hypothesis was originally proposed by \citet{Shibata82}, who used pressure gradients in hydrodynamic simulations to drive waves into the upper solar atmosphere. Two-dimensional MHD simulations of magnetic reconnection in the lower solar atmosphere have confirmed the formation of such shocks following magnetic reconnection (see, for example, \citealt{Kayshap13, Takasao13}). Some observations have also found evidence for the occurrence of these processes in the solar atmosphere (\citealt{dePontieu04, Tziotziou05, Heesu14}). Importantly, if this were the case, then a time lag (of the order minutes) should be present between the detection of reconnection processes in the photosphere and the occurrence of a surge, as the slow-wave would have to propagate from the reconnection site into the transition region.

\begin{figure*}
\includegraphics[width=0.95\textwidth]{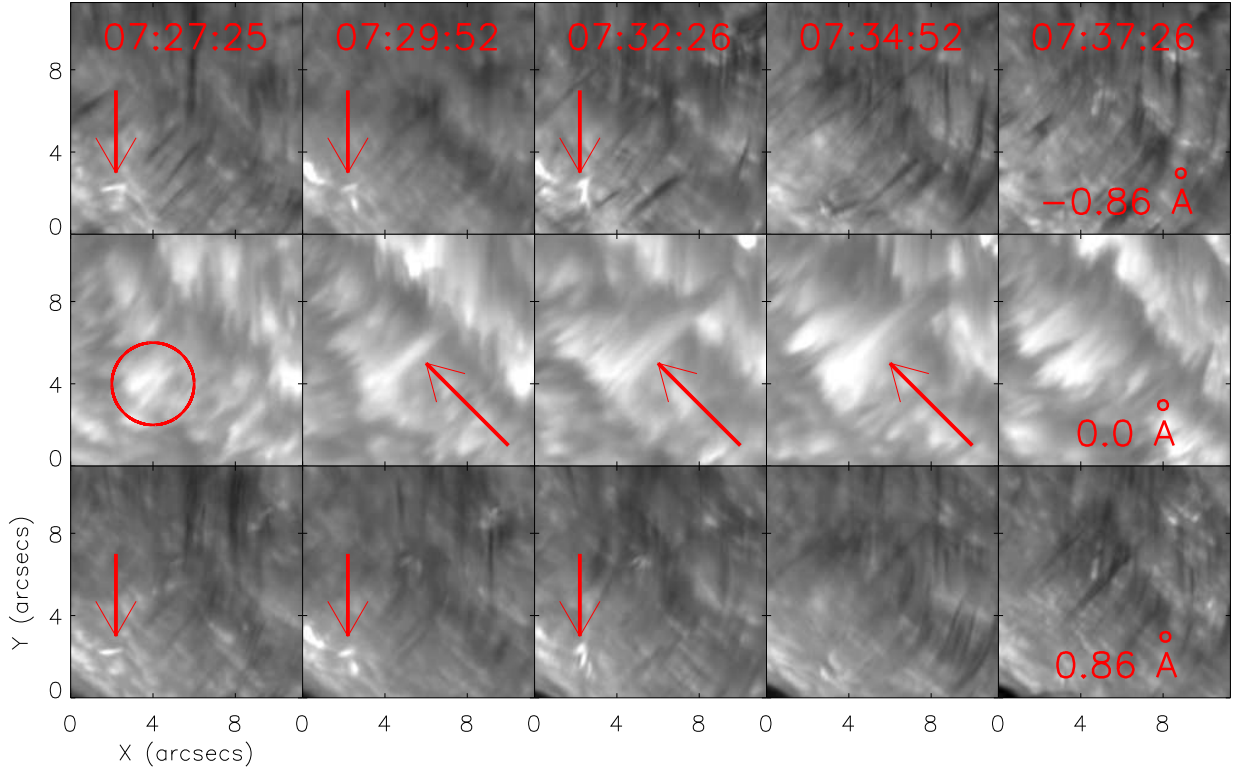}
\caption{The evolution of Event A through its lifetime sampled at three positions within the H$\alpha$ line profile. The top row plots the H$\alpha$ blue wing ($-0.86$ \AA), the middle row plots the H$\alpha$ line core, and the bottom row plots the H$\alpha$ red wing ($+0.86$ \AA). The arrows overlaid on the blue and red wing images indicate the inverted-Y shaped jet detected in the lower atmosphere which is studied in detail in Sect.~\ref{InvertYSec}. The associated surge is indicated by the arrows in the H$\alpha$ line core panels. The red circle in the left-hand H$\alpha$ line core panel identifies a small bright patch detected prior to Surge A (discussed in Sect.~\ref{SurgeSec}). A movie corresponding to the evolution of this event is included in the online version of this article.}
\label{SurgeA}
\end{figure*}

In this article, we analyse two inverted-Y shaped jets observed in AR 11506 at the solar limb using high-resolution H$\alpha$ data in order to discern whether signatures of bi-directional flows are present. Sustained asymmetries in the H$\alpha$ line profiles within the foot-points of the inverted-Y shaped jets imply the presence of bi-directional reconnection out-flows, providing evidence that these events formed as a response to magnetic reconnection in the lower solar atmosphere. Additionally, both inverted-Y shaped jets appear to form at the foot-points of surges which appear three minutes after the apparent on-set of reconnection. We set out our work as follows: In Sect.~\ref{Observations} we introduce the data studied in this article; In Sect.~\ref{Results} we present our results; In Sect.~\ref{Discussion} we provide a discussion of our results before we draw our conclusions in Sect.~\ref{Conclusions}.

\section{Observations}
\label{Observations}

In this article, we use ground-based data collected using the CRisp Imaging SpectroPolarimeter (CRISP; \citealt{Scharmer06, Scharmer08}) at the Swedish $1$-m Solar Telescope (SST; \citealt{Scharmer03}).  These data sampled AR $11506$ between $07$:$15$:$09$ UT and $07$:$48$:$25$ UT on $21$st June $2012$ (approximate co-ordinates of $x_\mathrm{c}$=$893$\arcsecs\, $y_\mathrm{c}$=$-250$\arcsecs) and have a pixel scale of around $0.059$\arcsecs\ (equating to around $43$ km in the horizontal plane) and a cadence of $7.7$ s, following reduction using the Multi-Object Multi-Frame Blind Deconvolution (MOMFBD; \citealt{Noort05}) method. The reduction employed the standard CRISP reduction pipeline (discussed by \citealt{Rodriguez15}), including the post-MOMFBD correction for differential stretching (\citealt{Henriques12}). A total of  $35$ wavelength positions across the H$\alpha$ line profile (where $6562.8$ \AA\ is now used throughout the remainder of this article as a reference [$0$ \AA] wavelength) were observed in the range [$-2$ \AA, $+1.2$ \AA].

In order to supplement the CRISP data, we also used space-borne data sampled by the Solar Dynamics Observatory's Atmospheric Imaging Assembly (SDO/AIA; \citealt{Lemen12}). These data are analysed between $07$:$15$:$09$ UT and $08$:$19$:$48$ UT in order to include the full evolution of the second event discussed in this article. We use $1600$ \AA\ and $1700$ \AA\ data, with a cadence of $24$ s, to investigate whether brightenings were observed in the lower atmosphere and three EUV channels ($304$ \AA, $171$ \AA, and $211$ \AA), with a cadence of $12$ s, to deduce the response of the upper atmosphere. All of these data have a pixel scale of approximately $0.6$\arcsecs\ (corresponding to around $430$ km in the horizontal scale). The data analysis presented here was conducted, in part, using the CRISPEX tool (\citealt{Vissers12}). It should be noted that although a careful alignment between the SDO/AIA data and the CRISP data was completed to confirm that the ejections observed in the EUV channels corresponded to those detected in the ground-based data, the analysis presented here was conducted on the non-rotated data in order to limit errors induced by interpolating the images. In Fig.~\ref{SDOtime}, we plot a time-series of context images displaying the FOV analysed here as observed by the SDO/AIA $304$ \AA\ (top row), $171$ \AA\ (middle row), and $211$ \AA\ (bottom row) filters. The labelled white arrows in the top row indicate the locations of the two surges studied in Sect.~\ref{SurgeSec}. 

\begin{figure*}
\includegraphics[width=0.95\textwidth]{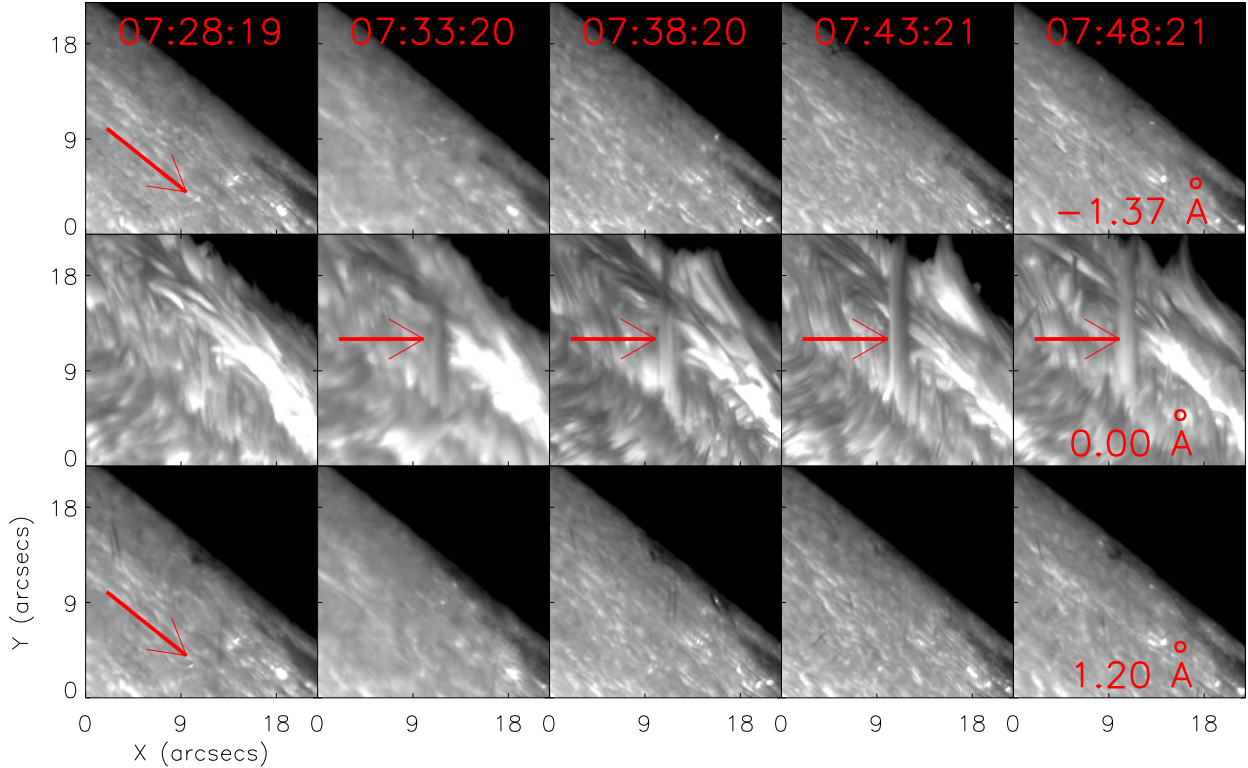}
\caption{Similar to Fig.~\ref{SurgeA} but for Event B. The H$\alpha$ line wing positions are now $-1.37$ \AA\ and $+1.20$ \AA\ to better highlight the inverted-Y shaped jet. The inverted-Y shaped jet (indicated by red arrows in the line wing images and studied in detail in Sect.~\ref{InvertYSec}) is faint in comparison to near-by EBs. The surge associated with this jet extends outside of the CRISP FOV (right-hand panel in the middle row). A movie corresponding to the evolution of this event is included in the online version of this article.}
\label{SurgeB}
\end{figure*}

\section{Results}
\label{Results}

\subsection{Properties of the inverted-Y shaped jets}
\label{InvertYSec}

We begin our analysis by investigating two inverted-Y shaped jets identified in the wings of the H$\alpha$ line (from around $\pm0.8$ \AA\ outwards). The evolution of the lower solar atmosphere during and following the first inverted-Y shaped jet studied here (where the inverted-Y shaped jet and apparently associated surge are known subsequently as `Event A' and `Surge A', respectively) was evident between $07$:$25$ UT and $07$:$33$ UT and is plotted in Fig.~\ref{SurgeA} for three positions in the H$\alpha$ line profile. Event A is clearly observable in the line wings (indicated by the arrows in the top and bottom rows) in the first three columns, however, no evidence of this structure is apparent in the H$\alpha$ line core. This inverted-Y shaped jet had an apparent vertical extent of approximately $1$\arcsecs\ and a foot-point separation of around $1$\arcsec. These values align well with the statistical properties of chromospheric anemone jets (as reported by \citealt{Nishizuka11}). Chromospheric anemone jets have been widely associated with magnetic reconnection, both observationally (as discussed by, {\it e.g.}, \citealt{Shibata07, Morita10, Singh11}) and numerically (\citealt{Yang13}). A similar plot is included for the second inverted-Y shaped jet studied here (`Event B' and `Surge B') in Fig.~\ref{SurgeB}. Event B was visible between approximately $07$:$28$ UT and $07$:$34$ UT and had similar spatial properties to Event A. The evolution of these events is plotted in the online movies associated with this article.

\begin{figure*}
\includegraphics[width=0.98\textwidth,trim={1cm 0 0 0}]{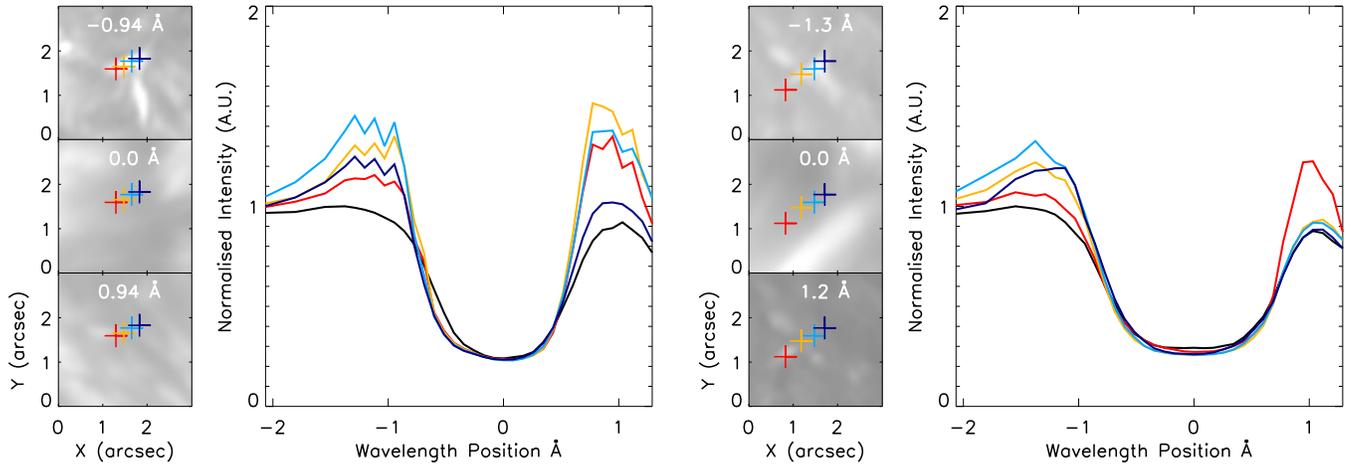}
\caption{(Left hand panels) Event A plotted in three wavelengths (specific wavelength positions indicated in the individual panels) in the H$\alpha$ line profile at $07$:$30$:$53$ UT. The crosses indicate pixels analysed along one leg of the inverted-Y shaped jet. The spectral profiles for the background (black line) and the four positions along the length of one leg of the inverted-Y shaped jet analysed here are plotted normalised against the background intensity. A clear transition from red- to blue-shifted profiles occurs from the bottom to the top of the leg. (Right hand panels) Same but for Event B. Again, a clear asymmetry in the profiles (from blue to red) can be observed along the leg of the jet.}
\label{Profiles}
\end{figure*}

In order to further investigate these inverted-Y shaped jets, we analysed the H$\alpha$ line profiles within their foot-points. Cartoon models of such jets imply that signatures of bi-directional flows (e.g., asymmetric line profiles) would be expected along the length of the jets (examples are presented in, {\it e.g.}, \citealt{Shibata07, Singh11}). These models suggest that such bi-directional flows are caused by magnetic reconnection which occurs within one leg of the jet. Although the associated asymmetries would perhaps be more prominent closer to disk centre due to the expected vertical nature of the magnetic reconnection, one would still expect some signature at the limb as it is unlikely that any reconnection out-flows would be purely in the plane-of-sky. In the left-hand panels of Fig.~\ref{Profiles}, we plot Event A at three positions within the H$\alpha$ line profile and the normalised line profiles at four points along the leg of the inverted-Y shaped jet. The four spatial positions at which the line profiles were calculated are indicated by coloured crosses on each image. The right-hand panels plot the same information for Event B.

At the base of Event A, strong asymmetries are detected in the line profiles, with larger intensity enhancements being measured in the red wing of the line when compared to the blue wing. These red-shifts (red and orange profiles in Fig.~\ref{Profiles}) are consistent with material flowing away from the observer. Symmetric profiles are then observed at the centre of the leg (light blue line in Fig.~\ref{Profiles}) before strong asymmetries manifesting as enhancements in the blue wing intensities (e.g., blue-shifted profiles) are measured at the top of the leg (purple line in Fig.~\ref{Profiles}), indicative of material moving towards the observer. Similar results are found for Event B, however, these, are perhaps more striking with the purple and red profiles displaying $40$ \% increases in intensity in one wing only. It should be noted that the asymmetries extend out to $-2$ \AA\ (as can be seen in Fig.~\ref{Profiles}), beyond the typical spectral extent of RBEs (\citealt{Sekse12}). These results suggest that spatially resolved bi-directional flows may be present within the foot-point of the jet, analogous to the cartoon models of chromospheric anemone jets (see \citealt{Shibata07, Singh11}). We stress that the identification of bi-directional flows within the legs of the inverted-Y shaped jets at the foot-points of the surges do not correspond to the up- and down-flows identified within surges themselves (e.g., \citealt{Brooks07, Madjarska09}).

\begin{figure*}
\includegraphics[scale=0.39]{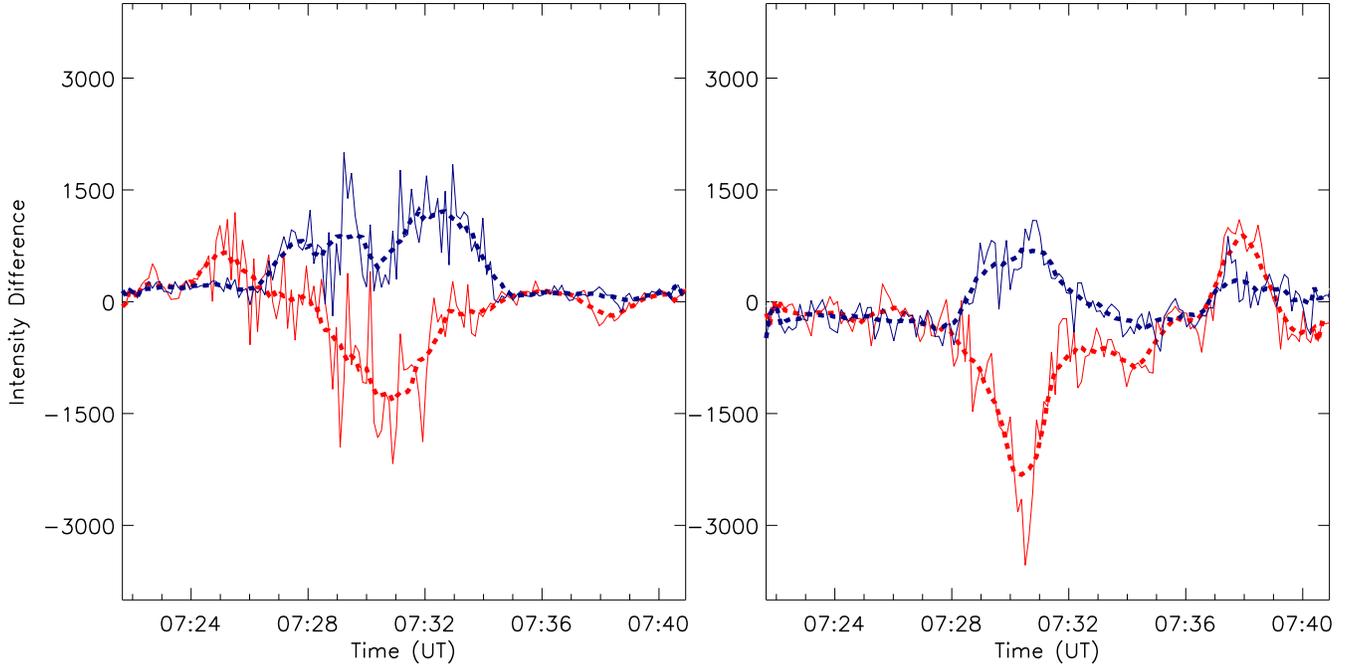}
\caption{The intensity difference between the blue wing and the red wing ($Int[-0.946$ \AA$]$-$Int[0.946$ \AA$]$) through time for two pixels at the top (purple line) and bottom (red line) of the inverted-Y shaped jets. Event A is plotted in the left-hand panel and Event B is plotted in the right-hand panel. The colours correspond to pixels denoted by the same coloured crosses plotted in Fig.~\ref{Profiles}. The over-laid dashed lines are running difference plots with smoothing applied over 10 frames (i.e., $77$ s). These plots clearly display that the asymmetries within the inverted-Y shaped jets are sustained for around $8$ minutes for Event A and $6$ minutes for Event B.}
\label{Dopp}
\end{figure*}

In order to better understand whether these asymmetries originate from bi-directional flows within the inverted-Y shaped jets themselves or are caused by line-of-sight effects, we conduct an analysis of the temporal evolution of the H$\alpha$ line profiles. Specifically, it is important to examine whether these asymmetries could be caused by the supposition of EB-like symmetric wing emission profiles with asymmetric wing absorption profiles caused by motion of material in the foreground (e.g. Rapid Blue Excursions; RBEs). In Fig.~\ref{Dopp}, we plot the difference in intensity between the red and blue wings ($\pm0.946$ \AA) through time for the purple and red pixels plotted in Fig.~\ref{Profiles}. For both Event A and Event B, the asymmetries are sustained throughout the lifetimes of the events which are longer than the average lifetimes of RBEs (see, for example, \citealt{Sekse12} where typical RBEs were found to be detectable for less than 90 s). As the lifetimes of these inverted-Y shaped jets is larger than those associated with RBEs, we can refute the hypothesis that these asymmetries could be caused by the supposition of symmetric emission and asymmetric absorption profiles.

The strong asymmetries within the individual line profiles plotted in Fig.~\ref{Profiles} (particularly evident in the red and purple lines in the right-hand panel where one wing is in emission and one wing is at the background intensity level) distinguish these inverted-Y shaped jets from both EBs (see \citealt{Nelson15} for an analysis of EBs in the same dataset) and Quiet-Sun Ellerman-like Brightenings (QSEBs; as discussed by \citealt{Rouppe16, Nelson17}). The relatively small intensity enhancements (only $140$ \% of the background intensity) of the inverted-Y shaped jets are also lower than the thresholds used to identify EBs in the modern literature (see, for example, \citealt{Vissers15, Nelson15}). To further investigate the  chromospheric anemone jets, we searched for signatures of these events in the SDO/AIA $1600$ \AA\ and $1700$ \AA\ channels. No increased emission was detected in either filter during the occurrence of the inverted-Y shaped jets which, again, distinguishing these events from EBs. 

As the inverted-Y shaped jets brighten in the wings of the H$\alpha$ line profile, and not in the line core, it is likely that the magnetic reconnection which drives these events takes place low down in the solar atmosphere, potentially in the photosphere (for a discussion of small-scale reconnection events in the lower solar atmosphere see \citealt{Young18}). The models of \citet{Takasao13} suggest that reconnection at such heights (i.e., below the $\beta$=$1$ layer) would generate slow magneto-acoustic wave pulses which would shock in the upper chromosphere to lift the transition region to greater heights, which we could observe as a surge. We examine potential links between these inverted-Y shaped jets and surges in the following sub-section.

\subsection{Potential links to surges}
\label{SurgeSec}

Both inverted-Y shaped jets studied in this article appeared to have some links to surges observed in the H$\alpha$ line core. For Event A, a surge was observed to form at the same apparant spatial location a few minutes after the on-set of the inverted-Y shaped jet. Initially, a small brightening (indicated by the red circle on the first panel of the middle row in Fig.~\ref{SurgeA}) was observed in the H$\alpha$ line core at around $07$:$27$:$25$ UT before a surge extended out into the upper atmosphere from this location. By $07$:$38$:$13$ UT the surge had receeded back to the chromospheric canopy. The evolution of Surge A through time in the H$\alpha$ line core can be seen in the middle row of Fig.~\ref{SurgeA}. Surge A evolved in two apparently distinct phases, specifically the initial rise phase and the subsequent descending phase. At the beginning of the rise phase, a small bright patch was present in the H$\alpha$ line core (enclosed in the red circle in the first panel of the middle row in Fig.~\ref{SurgeA}) at the foot-point of the event. This bright patch had a length of around $2$\arcsecs, a width of approximately $1$\arcsecs, a lifetime of around $1$ minute, and faded as Surge A began to increase in length. Throughout the rise phase, Surge A manifested as a thin, collimated structure (indicated by the red arrows in the second and third panels of the middle row of Fig.~\ref{SurgeA}), with a width of around $2.2$\arcsecs. The peak length of the structure was around $8.1$\arcsecs, which was reached at $07$:$32$:$41$ UT. This makes Surge A a shorter than average surges (\citealt{Roy73b}). 

After Surge A had reached its peak length and began to receed, it began to expand radially. This behaviour is similar to the expected evolution of surges (or jets) when magnetic reconnection takes place in the photosphere or lower chromosphere, according to the simulations of \citet{Takasao13}. In this scenario, slow wave pulses, formed due to the occurrence of magnetic reconnection in the lower solar atmosphere, shock in the transition region leading to the ejection of chromospheric material, initially along the magnetic field lines, which we observe as a surge. The peak width of Surge A was approximately $5$\arcsecs\ (see the fourth panel of the middle row in Fig.~\ref{SurgeA}) before it disappeared from view entirely (fifth panel of the H$\alpha$ line core row of Fig.~\ref{SurgeA}) as the surge material fell back under gravity to normal heights.

Surge B, the surge apparently associated with Event B, was first observable in the H$\alpha$ line core at $07$:$30$:$38$ UT and lived beyond the end of the CRISP observations. As, here, the surge is resolvable in a range of SDO/AIA filters, the lack of high-resolution ground-based data sampling the descending phase of the surge does not limit our ability to infer information about its evolution. Through analysis of the SDO/AIA $304$ \AA\ channel, we were able to identify that Surge B receeded back to pre-surge heights at around $07$:$54$:$20$ UT, giving it a total lifetime of around $24$ minutes. As can be seen in the centre row of Fig.~\ref{SurgeB}, the surge extended out of the CRISP field-of-view (FOV) meaning the spatial properties of the structure were also measured using the SDO/AIA $304$ \AA\ channel. The peak length of Surge B was $23$\arcsecs\ and its width varied between $2.5$-$3.5$\arcsecs\ through time. No expansion of the surge material was detected during the descending phase.

\begin{figure}
\includegraphics[scale=0.31,trim={0 0 0 0}]{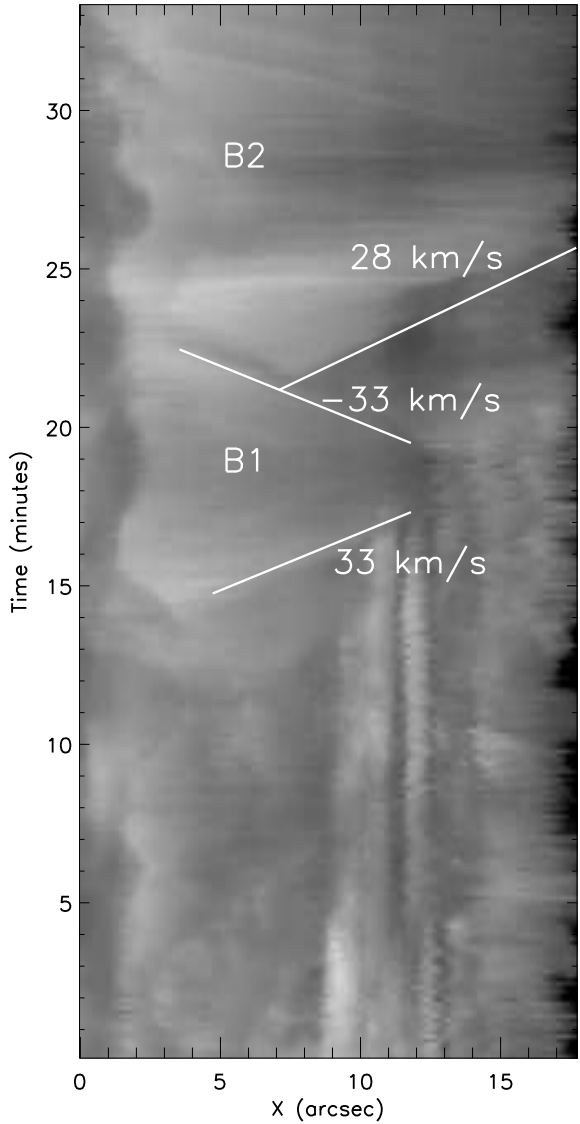}
\caption{Distance-time plot displaying the evolution of the length of Surge B through time plotted with a natural log intensity scale. The black line in the right-hand panels of Fig.~\ref{Invert} indicate the pixels used to construct this plot. Time zero corresponds to $07$:$15$:$09$ UT (i.e., the beginning of the observational time-series). The white lines indicate the position of the tops of the surge material through time for ease of the reader (and thus represent the apparent velocity). The recurrent nature of Surge B is evident, with the two repetitions being indicated by the labels `B1' and `B2'. The apparent upward (positive) and downward (negative) velocities of the surges are indicated by the labelled white lines over-laid on the plot.}
\label{SurgeTimDis}
\end{figure}

\begin{figure*}
\includegraphics[scale=0.45,trim={1cm 0 0 0}]{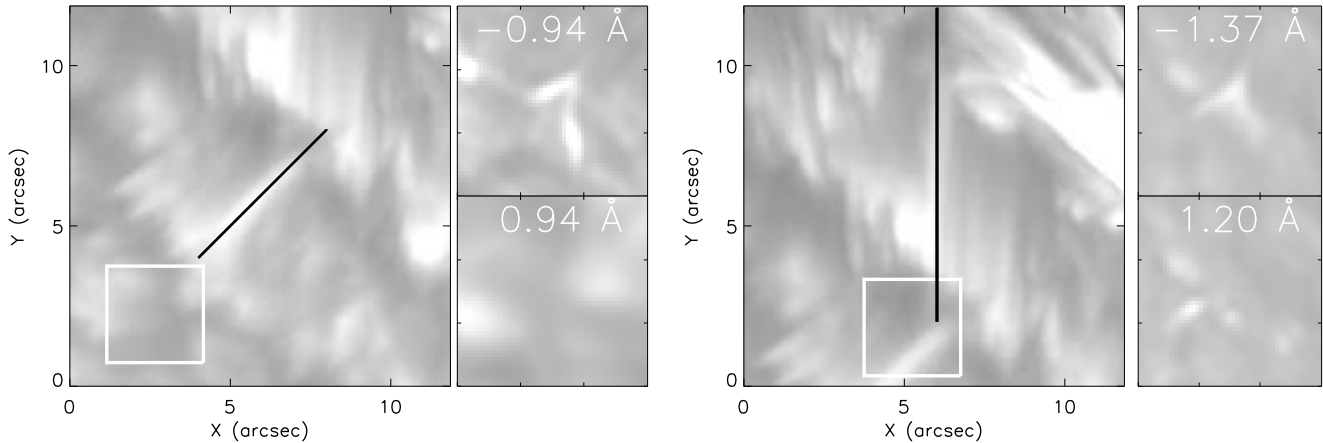}
\caption{(Left hand panels) The chromospheric H$\alpha$ line core (large panel) and photospheric H$\alpha$ line wing (smaller panels) signatures of Surge A and its foot-points at $07$:$30$:$38$ UT. The black line on the left hand panel indicates the positioning and orientation of the surge. The white boxes indicates the FOV plotted in the smaller panels. (Right hand panels) Same but for Surge B. The black line overlaid on the H$\alpha$ line core image indicates the axis of the slit used to construct the distance-time diagram plotted in Fig.~\ref{SurgeTimDis}. The inverted-Y shaped jets are clearly evident at the foot-point of both surges in the blue wing of the H$\alpha$ line in this time-step.}
\label{Invert}
\end{figure*}

In terms of temporal evolution, Surge B was slightly more complex than Surge A. In Fig.~\ref{SurgeTimDis}, we plot a time-distance diagram constructed by sampling the H$\alpha$ line core intensity along the length of the surge. The exact positioning of the slit used to construct the diagram is indicated by the black line in the right-hand panels of Fig.~\ref{Invert}. The event classified as Surge B here appears to be made-up of two successive surges which occurred at the same location (labelled as `B1' and `B2') around six minutes apart. The first ejection contained within the surge was relatively short, with a length of around $8$\arcsecs; however, the second ejection reached much higher into the atmosphere, to heights of around $23$\arcsecs\ (measured using the $304$ \AA\ channel from SDO/AIA). Both surges extend with an apparent upward (plane-of-sky) velocity of around $30$ km s$^{-1}$ ($33$ km s$^{-1}$ and $28$ km s$^{-1}$ for surges B1 and B2, respectively), slower than the surges discussed by \citet{Kayshap13}. Surge B1 also receeded at a similar apparent velocity (see Fig.~\ref{SurgeTimDis}). The occurrence of multiple ejections along the same trajectory implies the occurrence of a repetitive driver, in this case thought to be magnetic reconnection in the lower solar atmosphere. Such repetition has been observed in EBs (\citealt{Vissers15}), UV bursts (\citealt{Nelson16}) and explosive events (\citealt{Chae98}).

When observed with the SDO/AIA EUV filters, the main bodies of both surges appeared in aborption throughout their lifetimes, and no increased emission was detected at their foot-points or tips during their rise or descent phases (see Fig.~\ref{SDOtime}). The fact that both surges appeared in absorption implies the presence of higher levels of neutral Hydrogen along the line-of-sight (\citealt{Williams13}). Additionally, no bright material was present at the tips of these events, unlike the wave-driven events recently studied by \citet{Reid18}, suggesting no significant heating was occurring. These observations match well with the hypothesis that relatively cool, chromospheric material is lifted into the upper atmosphere by slow shock pulses at the transition region (although the SDO/AIA data themselves do not rule out other scenarios). Interestingly, Surge B appears to be well aligned to a coronal loop arcade which can be identified in $171$ \AA\ images (see panels three and four of the middle row of Fig.~\ref{SDOtime}). The material contained within Surge B is seemingly confined within the lower portions of the loop during both its rise and descent phases, however, direct alignment is difficult due to the relatively low spatial resolution of the SDO/AIA data.

In order to further analyse both surges, we investigated whether any rotational motions were evident during their evolutions or whether the inclination angle of the surges could be inferred. Previous research has indicated that the rotation of solar jets can be related to the degree of twist within the reconnection site at their foot-points (see, for example, \citealt{Liu18}). Therefore, a lack of observable twist within the surges could add support to the hypothesis that these events formed as a response to upwardly propagating slow shock pulses, rather than the release of free magnetic energy stored within twisted magnetic field lines. Doppler maps were constructed through single Gaussian fitting to the observed H$\alpha$ line profiles, in the wavelength range [$-1.032$ \AA, $+1.032$ \AA], at each pixel in the FOV. Velocities were estimated by comparing the position of the centre of the fitted Gaussian to $6562.8$ \AA, the assumed rest wavelength of H$\alpha$. This analysis identified no significant (greater than $\pm1$ km s$^{-1}$) line-of-sight velocities co-spatial to either event in terms of either rotation or inclination.

In Fig.~\ref{Invert}, we plot the locations of the two surges analysed here in the chromospheric H$\alpha$ line core (larger panels) in comparison to the inverted-shaped jets in the H$\alpha$ line wings (smaller panels). All frames are plotted at approximately $07$:$30$:$38$ UT. The black lines in the H$\alpha$ line core images indicate the axis of the surges and the white boxes outline the FOV plotted in the line wings at their foot-points in the smaller panels. The specific line wing positions are denoted on each individual panel. The inverted-Y shaped jets are both apparent at the foot-points of the surges. It should be noted that for Surge B, the inverted-Y shaped jet is only evident prior to B1. Only a small brightening in the blue wing of the H$\alpha$ line with an indistinct shape is detectable prior to B2, however, this could be due to a reduction in the seeing level during this time.

As can be seen in Fig.~\ref{SurgeA} and Fig.~\ref{SurgeB}, the inverted-Y shaped jets occured prior to and during the initial extension phases of both surges. Both inverted-Y shaped jets are first observable around three minutes prior to the formation of the surges. By measuring the height difference between the reconnection site (where the slow mode wave pulse is conjectured to be excited) and the transition region (where the slow mode shock pulse theoretically lifts the surge material), we would be able to calculate the propagation speed of the slow wave pulse; however, as this height difference is not known and cannot be measured, here we instead use potential height differences to estimate representative speeds of the propagating wave pulse. If we assume representative height differences of $1000$ km, $1500$ km, and $2000$ km, we are able to estimate potential slow wave propagation speeds of $5.5$ km s$^{-1}$, $8.3$ km s$^{-1}$, and $11.1$ km s$^{-1}$. These values are in line with the expected propagation speed of slow MHD waves in the lower solar atmosphere. It is possible, therefore, that these surges are formed as a response to magnetic reconnection in the lower solar atmosphere.

\section{Discussion}
\label{Discussion}

In this article, we have presented an analysis of two inverted-Y shaped jets detected in the wings of H$\alpha$ line scans sampled by the CRISP instrument (see Fig.~\ref{SurgeA} and Fig.~\ref{SurgeB}). These inverted-Y shaped jets had similar properties (lifetimes, lengths, foot-point separations) to chromospheric anemone jets (see, for example, \citealt{Shibata07, Nishizuka11}) as well as other inverted-Y shaped jets (e.g., \citealt{Tian18}). The intensity enhancements measured at the locations of these jets are relatively faint, with a maximum intensity enhancement of around $140$ \% (Fig.~\ref{Profiles}). This is well below the $150$ \%\ threshold value often used for EBs (e.g., \citealt{Nelson15}).

Strong asymmetries in the H$\alpha$ line profiles were measured in the legs of both inverted-Y jets (often with intensities $20$-$30$ \% higher in one wing over another). These asymmetries progressed from blue to red along the length of the inverted-Y shaped jets implying that bi-directional flows were occurring. Such flows would be expected if these inverted-Y shaped jets were formed due to magnetic reconnection (discussed by, for example, \citealt{Shibata07}). These asymmetries last for the entire lifetime of the inverted-Y shaped jets ($9$ minutes and $6$ minutes for Event A and Event B, respectively) indicating that they are physical in nature rather than due to a supposition of symmetric EB-like profiles and asymmetric RBE-like profiles (which would have much shorter lifetimes). In a manner similar to EBs, no brightening signature is observed in the H$\alpha$ line core during these jets implying that magnetic reconnection took place low down in the atmosphere, potentially in the photosphere. 

Both inverted-Y shaped jets appeared to form at the foot-points of chromospheric surges. The surge associated with Event A was a relatively short event, with a length of $8$\arcsecs, and had a lifetime of just under $11$ minutes. After this event had reached its peak height and begun to contract, it started to expand radially in a manner similar to the surges presented in the simulations of \citet{Takasao13}. The surge associated with Event B was longer, with a maximum length of $23$\arcsecs, however, it appeared to be comprised of two consecutive surges at the same location. Both surges appeared in absorption in SDO/AIA EUV data implying the presence of H bound-free absorption (i.e., an increase in the neutral Hydrogen density, consist with the hypothesis of chromospheric material being lifted higher into the atmosphere) as was discussed by \citet{Williams13}. A localised bright region observed at the foot-point of Surge A in the H$\alpha$ line core could be indicative of some heating at the chromospheric or transition region layers, however, no brightening was evident in SDO/AIA data. Interestingly, a three minute delay was also observed between the detection of the inverted-Y shaped jets and the formation of the surges potentially corresponding to the time required for the slow-mode waves to propagate from the reconnection site to the chromosphere.

\section{Conclusions}
\label{Conclusions}

Overall, our results imply that these two inverted-Y shaped jets are formed as a response to magnetic reconnection in the lower solar atmosphere. One of the main advances of this work over previous works is the detection of bi-directional flows, which we were able to measure along the legs of the inverted-Y shaped jets using transition of blue to red asymmetries. Such flows would only be detectable in extremely high-resolution data such as those sampled by the CRISP instrument. Surges appear to form at the same locations as the inverted-Y shaped jets potentially further hinting to the formation of magnetic reconnection in the solar photosphere. However, we should note that our small sample size does not conclusively link inverted-Y shaped jets to surges meaning future work must be conducted. A larger statistical sample should be conducted in the future to provide further evidence about links between these two phenomena.

Future work should aim to study a larger statistical sample of inverted-Y shaped jets to identify how frequently such features form co-spatial to surges. Additionally, it would be interesting to study how the inclination angles of any associated surges relates to the flow patterns detected within the inverted-Y shaped jets. Such work would require extremely high-resolution spectroscopic data acquired during a period of good seeing due to the small-spatial scales and low intensity enhancements of these events. DKIST should offer excellent data in this regard.

\acknowledgements
We thank the UK Science and Technology Facilities Council (STFC; Grant numbers: ST/M000826/1 and ST/P000304/1) for the support received to conduct this research. The Swedish $1$-m Solar Telescope is operated on the island of La Palma by the Institute for Solar Physics of Stockholm University in the Spanish Observatorio del Roque de los Muchachos of the Instituto de Astrofísica de Canarias. The Institute for Solar Physics is supported by a grant for research infrastructures of national importance from the Swedish Research Council (registration number 2017-00625). We also thank Dr. E. M. Scullion for help with data reduction. SDO data are courtesy of NASA/SDO and the AIA science team.

\bibliographystyle{apj}
\nocite{*}
\bibliography{Surge_Formation}

\end{document}